\documentclass[12pt,titlepage]{article}
\begin{document}

\title{BOUND STATES OF THE DIRAC EQUATION FOR A CLASS OF EFFECTIVE QUADRATIC PLUS
INVERSELY QUADRATIC POTENTIALS}
\date{}
\author{Antonio S. de Castro \\
\\
UNESP - Campus de Guaratinguet\'{a}\\
Departamento de F\'{\i}sica e Qu\'{\i}mica\\
Caixa Postal 205\\
12516-410 Guaratinguet\'{a} SP - Brasil\\
\\
E-mail address: castro@feg.unesp.br (A.S. de Castro)}
\maketitle

\begin{abstract}
The Dirac equation is exactly solved for a pseudoscalar linear plus
Coulomb-like potential in a two-dimensional world. This sort of potential
gives rise to an effective quadratic plus inversely quadratic potential in a
Sturm-Liouville problem, regardless the sign of the parameter of the linear
potential, in sharp contrast with the Schr\"{o}dinger case. The generalized
Dirac oscillator already analyzed in a previous work is obtained as a
particular case.
\end{abstract}

The Schr\"{o}dinger equation with a quadratic plus inversely quadratic
potential, known as singular oscillator, is an exactly solvable problem \cite
{lan}-\cite{pr} which works for constructing solvable models of $N$
interacting bodies \cite{cal1}-\cite{cal2} as well as a basis for
perturbative expansions and variational analyses for spiked harmonic
oscillators \cite{hal1}-\cite{hal7}. Generalizations for finite-difference
relativistic quantum mechanics \cite{nag} as well as for time-dependent
parameters in the nonrelativistic version have also been considered \cite
{cam}-\cite{dod}. In the present paper we approach the time-independent
Dirac equation with a conserving-parity pseudoscalar potential given by a
linear plus a Coulomb-like potential. This sort of potential gives rise to
an effective quadratic plus inversely quadratic potential in a
Sturm-Liouville problem, regardless the sign of the parameter of the linear
potential. Therefore, this sort of problem is exactly solvable in the Dirac
equation. The generalized Dirac oscillator \cite{asc4} is obtained as a
particular case. In addition to its importance as a new solution of the
Dirac equation this problem might be relevant to studies of confinement of
neutral fermions by a linear plus inversely linear electric field.

The two-dimensional Dirac equation can be obtained from the
four-dimen\-sional one with the mixture of spherically symmetric scalar,
vector and anomalous magnetic-like (tensor) interactions. If we limit the
fermion to move in the $x$-direction ($p_{y}=p_{z}=0$) the four-dimensional
Dirac equation decomposes into two equivalent two-dimensional equations with
2-component spinors and 2$\times $2 matrices \cite{str}. Then, there results
that the scalar and vector interactions preserve their Lorentz structures
whereas the anomalous magnetic interaction turns out to be a pseudoscalar
interaction. Furthermore, in the 1+1 world there is no angular momentum so
that the spin is absent. Therefore, the 1+1 dimensional Dirac equation allow
us to explore the physical consequences of the negative-energy states in a
mathematically simpler and more physically transparent way. The confinement
of fermions by a pure \-con\-ser\-ving-parity pseudoscalar double-step
potential \cite{asc2} and their scattering by a pure
non\-con\-ser\-ving-parity pseudoscalar step potential \cite{asc3} have
already been analyzed in the literature providing the opportunity to find
some quite interesting results. Indeed, the two-dimensional version of the
anomalous magnetic-like interaction linear in the radial coordinate,
christened by Moshinsky and Szczepaniak \cite{ms} as Dirac oscillator, has
also received attention. Nogami and Toyama \cite{nt}, Toyama \textit{et al.} 
\cite{tplus} and Toyama and Nogami \cite{tn} studied the behaviour of wave
packets under the influence of such a conserving-parity potential,
Szmytkowski and Gruchowski \cite{sg} proved the completeness of the
eigenfunctions and Pacheco \textit{et al.} \cite{pa} studied some
thermodynamics properties. More recently de Castro \cite{asc4} analyzed the
possibility of existence of confinement of fermions under the influence of
pseudoscalar power-law potentials, including the case of potentials
unbounded from below, and discussed in some detail the eigenvalues and
eigenfunctions for conserving- and nonconserving-parity linear potentials.

In the presence of a time-independent pseudoscalar potential the 1+1
dimensional time-independent Dirac equation for a fermion of rest mass $m$
reads

\begin{equation}
\left( c\alpha p+\beta mc^{2}+\beta \gamma ^{5}V_{p}\right) \psi =E\psi
\label{eq1a}
\end{equation}

\noindent where $E$ is the energy of the fermion, $c$ is the velocity of
light and $p$ is the momentum operator. $\alpha $ and $\beta $ are Hermitian
square matrices satisfying the relations $\alpha ^{2}=\beta ^{2}=1$, $%
\left\{ \alpha ,\beta \right\} =0$. From the last two relations it follows
that both $\alpha $ and $\beta $ are traceless and have eigenvalues equal to 
$\pm $1, so that one can conclude that $\alpha $ and $\beta $ are
even-dimensional matrices. One can choose the 2$\times $2 Pauli matrices
satisfying the same algebra as $\alpha $ and $\beta $, resulting in a
2-component spinor $\psi $. The positive definite function $|\psi |^{2}=\psi
^{\dagger }\psi $, satisfying a continuity equation, is interpreted as a
probability position density and its norm is a constant of motion. This
interpretation is completely satisfactory for single-particle states \cite
{tha}. It is worth to note that the Dirac equation is covariant under $%
x\rightarrow -x$ if $V(x)$ changes sign. This is because the parity operator 
$P=\exp (i\varepsilon )P_{0}\sigma _{3}$, where $\varepsilon $ is a constant
phase and $P_{0}$ changes $x$ into $-x$, changes sign of $\alpha $ and $%
\beta \gamma ^{5}$ but not of $\beta $. Using $\alpha =\sigma _{1}$ and $%
\beta =\sigma _{3}$, $\beta \gamma ^{5}=\sigma _{2}$ and provided that the
spinor is written in terms of the upper and the lower components 
\begin{equation}
\psi =\left( 
\begin{array}{c}
\psi _{+} \\ 
\psi _{-}
\end{array}
\right)  \label{eq8a}
\end{equation}

\noindent the Dirac equation decomposes into :

\begin{eqnarray}
-\left( E-mc^{2}\right) \psi _{+} &=&i\hbar c\psi _{-}^{\prime }+iV\psi _{-}
\nonumber \\
&&  \label{eq8b} \\
-\left( E+mc^{2}\right) \psi _{-} &=&i\hbar c\psi _{+}^{\prime }-iV\psi _{+}
\nonumber
\end{eqnarray}

\noindent where the prime denotes differentiation with respect to $x$. In
terms of $\psi _{+}$ and $\psi _{-}$ the spinor is normalized as $%
\int_{-\infty }^{+\infty }dx\left( |\psi _{+}|^{2}+|\psi _{-}|^{2}\right) =1$%
, so that $\psi _{+}$ and $\psi _{-}$ are square integrable functions. It is
clear from the pair of coupled first-order differential equations (\ref{eq8b}%
) that $\psi _{+}$ and $\psi _{-}$ have definite and opposite parities if
the Dirac equation is covariant under $x\rightarrow -x$. In the
nonrelativistic approximation (potential energies small compared to $mc^{2}$
and $E\approx mc^{2}$) Eq. (\ref{eq8b}) loses all the matrix structure and
becomes

\begin{equation}
\psi _{-}=\frac{p}{2mc}\;\psi _{+}  \label{eq8c}
\end{equation}

\begin{equation}
\left( -\frac{\hbar ^{2}}{2m}\frac{d^{2}}{dx^{2}}+V_{eff}^{+}\right) \psi
_{+}=\left( E-mc^{2}\right) \psi _{+}  \label{eq8d}
\end{equation}

\noindent where $V_{eff}^{+}=V^{2}/2mc^{2}+\hbar /2mcV^{\prime }$. Eq. (\ref
{eq8c}) shows that $\psi _{-}$ if of order $v/c<<1$ relative to $\psi _{+}$
and Eq. (\ref{eq8d}) shows that $\psi _{+}$ obeys the Schr\"{o}dinger
equation with the effective potential $V_{eff}^{+}$. It is noticeable that
this peculiar (pseudoscalar-) coupling results in the Schr\"{o}dinger
equation with an effective potential in the nonrelativistic limit, and not
with the original potential itself. The form in which the original potential
appears in the effective potential, the $V^{2}$ term, allows us to infer
that even a potential unbounded from below could be a confining potential.
This phenomenon is inconceivable if one starts with the original potential
in the nonrelativistic equation.

The coupling between the upper and the lower components of the Dirac spinor
can be formally eliminated when Eq. (\ref{eq8b}) is written as second-order
differential equations:

\begin{equation}
-\frac{\hbar ^{2}}{2m}\;\psi _{\pm }^{\prime \prime }+V_{eff}^{\pm }\;\psi
_{\pm }=E_{eff}\;\psi _{\pm }  \label{30}
\end{equation}

\noindent where 
\begin{eqnarray}
E_{eff} &=&\frac{E^{2}-m^{2}c^{4}}{2mc^{2}}  \label{30-1} \\
&&  \nonumber \\
V_{eff}^{\pm } &=&\frac{V^{2}}{2mc^{2}}\pm \frac{\hbar }{2mc}V^{\prime }
\label{30-2}
\end{eqnarray}

\bigskip \noindent \noindent \noindent \noindent These last results show
that the solution for this class of problem consists in searching for
bounded solutions for two Schr\"{o}dinger equations. It should not be
forgotten, though, that the equations for $\psi _{+}$ or $\psi _{-}$ are not
indeed independent because the effective eigenvalue, $E_{eff}$, appears in
both equations. Therefore, one has to search for bound-state solutions for $%
V_{eff}^{+}$ and $V_{eff}^{-}$ with a common eigenvalue.

Now let us consider a pseudoscalar potential in the form 
\begin{equation}
V=m\omega cx+\frac{\hbar cg}{x}  \label{30b}
\end{equation}
\noindent where $\omega $ and $g$ are real parameters. Then the effective
potential becomes the singular harmonic oscillator 
\begin{equation}
V_{eff}^{\pm }=Ax^{2}+\frac{B_{\pm }}{x^{2}}+C_{\pm }  \label{32}
\end{equation}
\noindent where 
\begin{eqnarray}
A &=&\frac{1}{2}m\omega ^{2}  \nonumber \\
&&  \nonumber \\
B_{\pm } &=&\frac{\hbar ^{2}g}{2m}\left( g\mp 1\right)  \label{32b} \\
&&  \nonumber \\
C_{\pm } &=&\hbar \omega \left( g\pm \frac{1}{2}\right)  \nonumber
\end{eqnarray}

\noindent It is worthwhile to note at this point that the singularity at $x=0
$ never menaces the fermion to collapse to the center \cite{lan} because in
any condition $B_{\pm }$ is never less than the critical value $B_{c}=-\hbar
^{2}/8m$. Furthermore, $\omega \ $must be different from zero, otherwise
there would be an everywhere repulsive effective potential as long as the
condition $B_{+}<0$ and $B_{-}<0$ is never satisfied simultaneously.
Therefore, the parameters of the effective potential with $\omega \neq 0$
fulfill the key conditions to furnish spectra purely discrete with infinite
sequence of eigenvalues, regardless the signs of $\omega $ and $g$. Note
also that the parameters of the effective potential are related in such a
manner that the change $\omega \rightarrow -\omega $ induces the change $%
V_{eff}^{\pm }\rightarrow V_{eff}^{\pm }-2C_{\pm }$ ($C_{\pm }\rightarrow
-C_{\pm }$) whereas under the change $g\rightarrow -g$ one has $V_{eff}^{\pm
}\rightarrow V_{eff}^{\mp }-2C_{\mp }$ ($B_{\pm }\rightarrow B_{\mp }$ and $%
C_{\pm }\rightarrow -C_{\mp }$). The combined transformation $\omega
\rightarrow -\omega $ and $g\rightarrow -g$ has as effect $V_{eff}^{\pm
}\rightarrow V_{eff}^{\mp }$, meaning that the effective potential for $\psi
_{+}$ transforms into the effective potential for $\psi _{-}$ and vice
versa. In Figure 1 is plotted the effective potential for some illustrative
values of $g$. From this Figure one can see that in the case $|g|>1$ both
components of the Dirac spinor are subject to the singular harmonic
oscillator, for $|g|=1$ one component is subject to the harmonic oscillator
while the other one is subject to the singular harmonic oscillator, for $%
|g|<1$ one component is subject to a bottomless potential well and the other
one is subject to the singular harmonic oscillator, in the case $g=0$ both
components are subject to harmonic oscillator potentials.

Defining

\negthinspace 
\begin{equation}
\xi =\frac{\sqrt{2mA}}{\hbar }\;x^{2}  \label{33}
\end{equation}

\noindent and using (\ref{30})-(\ref{30-2}) one obtains the equation

\begin{equation}
\xi \psi _{\pm }^{\prime \prime }+\frac{1}{2}\psi _{\pm }^{\prime }+\left[ -%
\frac{\xi }{4}-\frac{mB_{\pm }}{2\hbar ^{2}\xi }+\frac{1}{4\hbar }\sqrt{%
\frac{2m}{A}}\left( E_{eff}-C_{\pm }\right) \right] \psi _{\pm }=0
\label{34}
\end{equation}

\noindent Now the prime denotes differentiation with respect to $\xi $. The
normalizable asymptotic form of the solution as $\xi \rightarrow \infty $ is 
$e^{-\xi /2}$. As $\xi \rightarrow 0$, when the term $1/x^{2}$ dominates,
the regular solution behaves as $\xi ^{s/2}$, where $s$ is a nonnegative
solution of the algebraic equation

\begin{equation}
s(s-1)-2mB_{\pm }/\hbar ^{2}=0  \label{34a}
\end{equation}
\textit{viz.}

\begin{equation}
s=\frac{1}{2}\left( 1\pm \sqrt{1+\frac{8mB_{\pm }}{\hbar ^{2}}}\right) \geq 0
\label{34b}
\end{equation}

\noindent If $B_{\pm }>0$ there is just one possible value for $s$ (that
ones with the plus sign in front of the radical) and the same is true for $%
B_{\pm }=B_{c}$ when $s=1/2$, but for $B_{c}<B_{\pm }<0$ there are two
possible values for $s$ in the interval $0<s<1$. If the singular potential
is absent ($B_{\pm }=0$) then $s=0$ or $s=1$. The solution for all $\xi $
can be expressed as $\psi _{\pm }(\xi )=\xi ^{s/2}e^{-\xi /2}w(\xi )$, where 
$w$ is solution of the confluent hypergeometric equation \cite{abr}

\begin{equation}
\xi w^{\prime \prime }+(b-\xi )w^{\prime }-aw=0  \label{35}
\end{equation}

\noindent with

\begin{eqnarray}
a &=&\frac{b}{2}-\frac{1}{4\hbar }\sqrt{\frac{2m}{A}}\left( E_{eff}-C_{\pm
}\right)  \nonumber \\
&&  \label{36} \\
b &=&s+1/2  \nonumber
\end{eqnarray}

\noindent Then $w$ is expressed as $_{1\!\;}\!F_{1}(a,b,\xi )$ and in order
to furnish normalizable $\psi _{\pm }$, the confluent hypergeometric
function must be a polynomial. This demands that $a=-n$, where $n$ is a
nonnegative integer in such a way that $_{1\!\;}\!F_{1}(a,b,\xi )$ is
proportional to the associated Laguerre polynomial $L_{n}^{b-1}(\xi )$, a
polynomial of degree $n$. This requirement, combined with the top line of (%
\ref{36}), also implies into quantized effective eigenvalues:

\begin{equation}
E_{eff}=\left( 2n+s+\frac{1}{2}\right) \hbar |\omega |+C_{\pm },\qquad
n=0,1,2,\ldots  \label{37}
\end{equation}

\noindent with eigenfunctions given by

\begin{equation}
\psi _{\pm }(x)=N_{\pm }\;\xi ^{s/2}e_{\;}^{-\xi /2}\;L_{n}^{s-1/2}\left(
\xi \right)  \label{38}
\end{equation}

\smallskip

\noindent On the other hand, $s=0$ and $s=1$ for the case $B_{\pm }=0$ and
the associated Laguerre polynomial $L_{n}^{-1/2}(\xi )$ and $%
L_{n}^{+1/2}(\xi )$ are proportional to $H_{2n}\left( \sqrt{\xi }\right) $
and $\xi ^{-1/2}H_{2n+1}\left( \sqrt{\xi }\right) $, respectively \cite{abr}%
. Therefore, the solution for the harmonic oscillator can be succinctly
written in the customary form in terms of Hermite polynomials:

\begin{eqnarray}
E_{eff} &=&\left( n+\frac{1}{2}\right) \hbar |\omega |+C_{\pm },\qquad
n=0,1,2,\ldots  \label{41a} \\
&&  \nonumber \\
\psi _{\pm }(x) &=&N_{\pm }e^{-\xi /2}H_{n}\left( \sqrt{\xi }\right)
\label{41b}
\end{eqnarray}

\noindent Note that the behaviour of $\psi _{\pm }$ at very small $\xi $
implies into the Dirichlet boundary condition ($\psi _{\pm }(0)=0)$ for $%
s\neq 0$. This boundary condition is essential whenever $B_{\pm }\neq 0$,
nevertheless it also develops for $B_{\pm }=0$ when $s=1$ but not for $s=0$.
Since $V_{eff}^{\pm }$ is invariant under reflection through the origin ($%
x\rightarrow -x$), eigenfunctions with well-defined parities can be found.
However, the eigenfunctions expressed in terms of associated Laguerre
polynomials, due to the presence of the term $x^{s}$, are restricted to the
half-line $x>0$ in the event of $s$ is not an integer number. This drawback
can be circumvented by taking $x$ by $|x|$. Then, there still an even
eigenfunction such as in the case when $s$ is an even number. New
eigenfunctions can be found by taking symmetric and antisymmetric linear
combinations of the eigenfunctions expressed in terms of eigenfunctions
defined on the positive side of the $x$-axis. These new eigenfunctions
possess the same effective eigenvalue so that there is a two-fold
degeneracy. This degeneracy in a one-dimensional quantum-mechanical problem
is due to the fact that the eigenfunctions, whether they are either even or
odd, disappear at the origin. These questions do not occur when the
eigenfunctions are expressed in terms of Hermite polynomials because they
are well-behaved on the entire $x$-axis and have both even and odd parities
from the beginning.

The necessary conditions for confining fermions in the Dirac equation with
the potential (\ref{30b}) have been put forward. The formal analytical
solutions have also been obtained. Now we move on to consider a survey for
distinct cases in order to match the common effective eigenvalue. As we will
see this survey leads to additional restrictions on the solutions, including
constraints involving the nodal structure of the upper and lower components
of the Dirac spinor. The effective potentials for a few specific cases are
already plotted in Figure 1 and all the others can be obtained by making the
transformation $\omega \rightarrow -\omega $ and $g\rightarrow -g$, as
mentioned before.

\bigskip \bigskip

\noindent \textbf{Case A} \textbf{-- }$|g|>1$

This is the situation where $B_{\pm }>0$ and for $g>1$ ($g<-1$) $s=|g|$ for $%
V_{eff}^{+}$ ($V_{eff}^{-}$), and $s=1+|g|$ for $V_{eff}^{-}$ ($V_{eff}^{+}$%
). The upper and lower components of the Dirac spinor share the same
eigenvalue if the quantum numbers satisfy the relation

\begin{equation}
n_{-}=n+\frac{\varepsilon (\omega )-\varepsilon (g)}{2}  \label{42}
\end{equation}

\noindent where $\varepsilon (\omega )$ and $\varepsilon (g)$ stand for the
sign function and $n_{+}=n$ ($n_{-}$) is related to $\psi _{+}$ ($\psi _{-}$%
). The solutions are

\begin{eqnarray}
\psi _{+\varepsilon (g)} &=&N_{+\varepsilon (g)}\;\xi ^{|g|/2}e_{\;}^{-\xi
/2}\;L_{n}^{|g|-1/2}\left( \xi \right)  \nonumber \\
&&  \nonumber \\
\psi _{-\varepsilon (g)} &=&N_{-\varepsilon (g)}\;\xi ^{\left( 1+|g|\right)
/2}e_{\;}^{-\xi /2}\;L_{n+\frac{\varepsilon (\omega )-\varepsilon (g)}{2}%
}^{|g|+1/2}\left( \xi \right)  \label{43} \\
&&  \nonumber \\
E_{eff} &=&\left[ 2n+1-\frac{\varepsilon (g)}{2}+|g|+\varepsilon (\omega
)\left( g+\frac{1}{2}\right) \right] \hbar |\omega |  \nonumber
\end{eqnarray}

\noindent with

\begin{equation}
n\geq \frac{\varepsilon (g)-\varepsilon (\omega )}{2}  \label{44}
\end{equation}

\smallskip

\noindent and the further proviso that $n\geq 0$ if $\varepsilon (\omega
)>\varepsilon (g)$. Note that the subscripts $\pm \varepsilon (g)$ in $\psi $
is a sequel of the transformation $V_{eff}^{\pm }\rightarrow V_{eff}^{\mp
}-2C_{\mp }$ under $g\rightarrow -g$. Note also that the constraint (\ref{42}%
) asserts that the number of nodes of $\psi _{+}$ and $\psi _{-}$ just
differ by $\pm 1$.

\bigskip \bigskip

\noindent \textbf{Case B -- }$|g|=1$\textbf{\ }

In this situation $B_{+}=0$ ($B_{-}=0$) and $s=0$ or $1$ for $V_{eff}^{+}$ ($%
V_{eff}^{-}$), and $B_{-}>0$ ($B_{+}>0$) and $s=2$ for $V_{eff}^{-}$ ($%
V_{eff}^{+}$) when $g=+1$ ($g=-1$). It follows from (\ref{37}) and (\ref{41a}%
) that the quantum numbers satisfy $n_{-\varepsilon (g)}=\left\{
n_{+\varepsilon (g)}-\left[ 2-\varepsilon (g)\varepsilon (\omega )\right]
\right\} /2$. Note that $n_{-\varepsilon (g)}$\noindent \ is an integer
number if and only if $n_{+\varepsilon (g)}$ is an odd number. Making the
substitution $n\rightarrow 2n+1$ one can get the same constraint as in Case
A, Eq. (\ref{42}), and the solutions can be written as

\begin{eqnarray}
\psi _{+\varepsilon (g)} &=&N_{+\varepsilon (g)}\;\xi ^{|g|/2}e_{\;}^{-\xi
/2}\;H_{2n+1}\left( \sqrt{\xi }\right)  \nonumber \\
&&  \nonumber \\
\psi _{-\varepsilon (g)} &=&N_{-\varepsilon (g)}\;\xi e_{\;}^{-\xi /2}\;L_{n+%
\frac{\varepsilon (\omega )-\varepsilon (g)}{2}}^{3/2}\left( \xi \right)
\label{46} \\
&&  \nonumber \\
E_{eff} &=&\left[ 2n+2-\frac{\varepsilon (g)}{2}+\varepsilon (\omega )\left(
g+\frac{1}{2}\right) \right] \hbar |\omega |  \nonumber
\end{eqnarray}

\noindent and the same remarks for the Case A involving $\psi _{\pm
\varepsilon (g)}$, as well as the nodal structure of the eigenfunctions,
apply in Case B.

\bigskip\bigskip

\noindent \textbf{Case C -- \noindent }$0<|g|<1$

Here the solutions break in two sets regarding the two distinct
possibilities of $s$ for $B_{+}<0$ ($B_{-}<0$) for $V_{eff}^{+}$ ($%
V_{eff}^{-}$) when $0<g<1$ ($-1<g<0$), \textit{viz. } $s=|g|$ or $1-|g|.$ On
the other side, $B_{-}>0$ ($B_{+}>0$) for $V_{eff}^{-}$ ($V_{eff}^{+}$) and $%
s=1+|g|$. For the first set, $s=|g|$ for $V_{eff}^{+}$ ($V_{eff}^{-}$), and
the constraint involving the quantum numbers as well as the solutions for
the effective eigenvalues and eigenfunctions are the very same as those ones
presented in the Case A. For the second set, $s=1-|g|$ for $V_{eff}^{+}$ ($%
V_{eff}^{-}$), and the quantum numbers satisfy $n_{-\varepsilon
(g)}=n_{\varepsilon (g)}-|g|+\varepsilon (g)\varepsilon (\omega )/2$.
\noindent This second set provides $n$ equal to an integer number on the
condition that $|g|=1/2$ and this very particular case is already built in
the first set.

\bigskip \bigskip

\noindent \textbf{Case D -- }$g=0$

This is the situation where $B_{\pm }=0$ and $s=0$ or $1$. The quantum
numbers satisfy

\begin{equation}
n_{-}=n+\varepsilon (\omega )  \label{51}
\end{equation}

\noindent and the solutions are

\begin{eqnarray}
\psi _{+} &=&N_{+}e^{-\xi /2}H_{n}\left( \sqrt{\xi }\right)  \nonumber \\
&&  \nonumber \\
\psi _{-} &=&N_{-}e^{-\xi /2}H_{n+\varepsilon (\omega )}\left( \sqrt{\xi }%
\right)  \label{52} \\
&&  \nonumber \\
E_{eff} &=&\left[ n+\frac{1+\varepsilon (\omega )}{2}\right] \hbar |\omega |
\nonumber
\end{eqnarray}

\noindent with

\begin{equation}
n\geq -\varepsilon (\omega )  \label{53}
\end{equation}

\smallskip

\noindent and the further proviso that $n\geq 0$ if $\varepsilon (\omega
)=+1 $.

\bigskip \bigskip

The preceding analyses shows that the effective eigenvalues are equally
spaced with a step given by $2\hbar |\omega |$ when at least one the
effective potentials is singular at the origin. It is remarkable that the
level stepping is independent of the sign and intensity of the parameter
responsible for the singularity of the potential. We also note that there is
a continuous transition from the Case A to the Case C, despite the
appearance of the Hermite polynomial in the Case B. It should be remembered,
though, that Hermite polynomials can be seen as particular cases of
associated Laguerre polynomials. When both effective potentials become
nonsingular ($g=0$) the step switches abruptly to $\hbar |\omega |$. There
is a clear phase transition when $g\rightarrow 0$ due the disappearance of
the singularity for both components of the Dirac spinor. In the limit as $%
g\rightarrow 0$ the Neumann boundary condition, in addition to the Dirichlet
boundary condition always present for $g\neq 0$, comes to the scene. This
occurrence permits the appearance of even Hermite polynomials and their
related eigenvalues, which intercalate among the pre-existent eigenvalues
related to odd Hermite polynomials. The appearance of even Hermite
polynomials makes $\psi (0)\neq 0$ and this boundary condition is never
permitted when the singular potential is present, even though $g$ can be
small. You might also understand the lack of such a smooth transition by
starting from a nonsingular potential ($g=0$), when the solution of the
problem involves even and odd Hermite polynomials, and then turning on the
singular potential as a perturbation of the $g=0$ potential. Now the
``perturbative singular potential'' by nature demands, if is  either
attractive or repulsive, that $\psi (0)=0$ so that it naturally kills the
solution involving even Hermite polynomials. Furthermore, there is no
degeneracy in the spectrum for $g=0$.

In all the circumstances, at least one of the effective potentials has a
well structure and the highest well governs the value of the zero-point
energy. Matching the formal solutions for a common effective eigenvalue has
imposed additional restrictions on the allowed eigenvalues and relations
between the number of nodes of the upper and lower components of the Dirac
spinor have been obtained. A sharp limitation occurred in the Case B ($|g|=1$%
) when all the solutions involving even Hermite polynomials have been
suppressed. In the Case D ($g=0$), a case which cannot be obtained as a
limit of the preceding ones, the solutions of the generalized Dirac
oscillator \cite{asc4} were obtained. The Dirac eigenvalues are obtained by
inserting the effective eigenvalues in (\ref{30-1}). \noindent \noindent One
should realize that the Dirac energy levels are symmetrical about $E=0$. It
means that the potential couples to the positive-energy component of the
spinor in the same way it couples to the negative-energy component. In other
words, this sort of potential couples to the mass of the fermion instead of
its charge so that there is no atmosphere for the spontaneous production of
particle-antiparticle pairs. No matter the intensity and sign of the
coupling parameters, the positive- and the negative-energy solutions never
meet. There is always an energy gap greater or equal to $2mc^{2}$, thus
there is no room for transitions from positive- to negative-energy
solutions. This all means that Klein\'{}s paradox never comes to the
scenario.

\smallskip Figures 2-5 illustrate the behaviour of $|\psi _{+}|^{2}$, $|\psi
_{-}|^{2}$ and $|\psi |^{2}=|\psi _{+}|^{2}+|\psi _{-}|^{2}$ for the
positive-energy solutions of the ground-state contemplating all the distinct
classes of effective potentials plotted in Figure 1. The relative
normalization of $\psi _{+}$ and $\psi _{-}$ is obtained by substituting the
solutions directly into the original first-order coupled equations (\ref
{eq8b}). Comparison of these Figures shows that $|\psi _{+}|$ is larger than 
$|\psi _{-}|$ (for $E>0$) and that the fermion tends to avoid the origin
more and more as $|g|$ increases. A numerical calculation of the uncertainty
in the position (with $m=\omega =c=\hbar =1$) furnishes $0.844,1.130,1.346$
and $1.528$ for $g$ equal to $0,1/2,1$ and $3/2$, respectively.

In conclusion, we have succeed in searching for Dirac bounded solutions for
the pseudoscalar potential $V=m\omega cx+\hbar cg/x$. The satisfactory
completion of this task has been possible because the methodology of
effective potentials has transmuted the question into Sturm-Liouville
problems with effective quadratic plus inversely quadratic potentials for
both components of the Dirac spinor. As stated in the first paragraph of
this work, the anomalous magnetic-like interaction in the four-dimensional
world turns into a pseudoscalar interaction in the two-dimensional world.
The anomalous magnetic interaction has the form $-i\mu \beta \vec{\alpha}.%
\vec{\nabla}\phi (r)$, where $\mu $ is the anomalous magnetic moment in
units of the Bohr magneton and $\phi $ is the electric potential, \textit{%
i.e.}, the time component of a vector potential \cite{tha}. Therefore,
besides its importance as new analytical solutions of a fundamental equation
in physics, the solutions obtained in this paper might be of relevance to
the confinement of neutral fermions in a four-dimensional world.

\bigskip

\noindent \textbf{Acknowledgments}

This work was supported in part by means of funds provided by CNPq and
FAPESP.

\smallskip \pagebreak

\bigskip \pagebreak

\noindent \textbf{Figure captions}

\medskip

\noindent Figure 1 -- Effective potentials for $V=m\omega cx+\hbar cg/x$
with $\omega >0$, $g\geq 0$. The solid lines are for $V_{eff}^{+}$, the
dashed lines are for $V_{eff}^{-}$. \quad a) $g=3/2$ ; \quad b) $g=1$ ;
\quad c) $g=1/2$ ; \quad d) $g=0$ \quad ($m=\omega =c=\hbar =1$).

\medskip

\noindent Figure 2 -- $|\psi _{+}|^{2}$ (full thin line), $|\psi _{-}|^{2}$
(dashed line), $|\psi _{+}|^{2}+|\psi _{-}|^{2}$ (full thick line),
corresponding to \emph{positive}-ground-state energy for the potential $%
V=m\omega cx+\hbar cg/x$ with $g=3/2$ and $m=\omega =c=\hbar =1$.

\medskip

\noindent Figure 3 -- The same as in Figure 2 with $g=1$.

\medskip

\noindent Figure 4 -- The same as in Figure 2 with $g=1/2$.

\medskip

\noindent Figure 5 -- The same as in Figure 2 with $g=0$.

\medskip

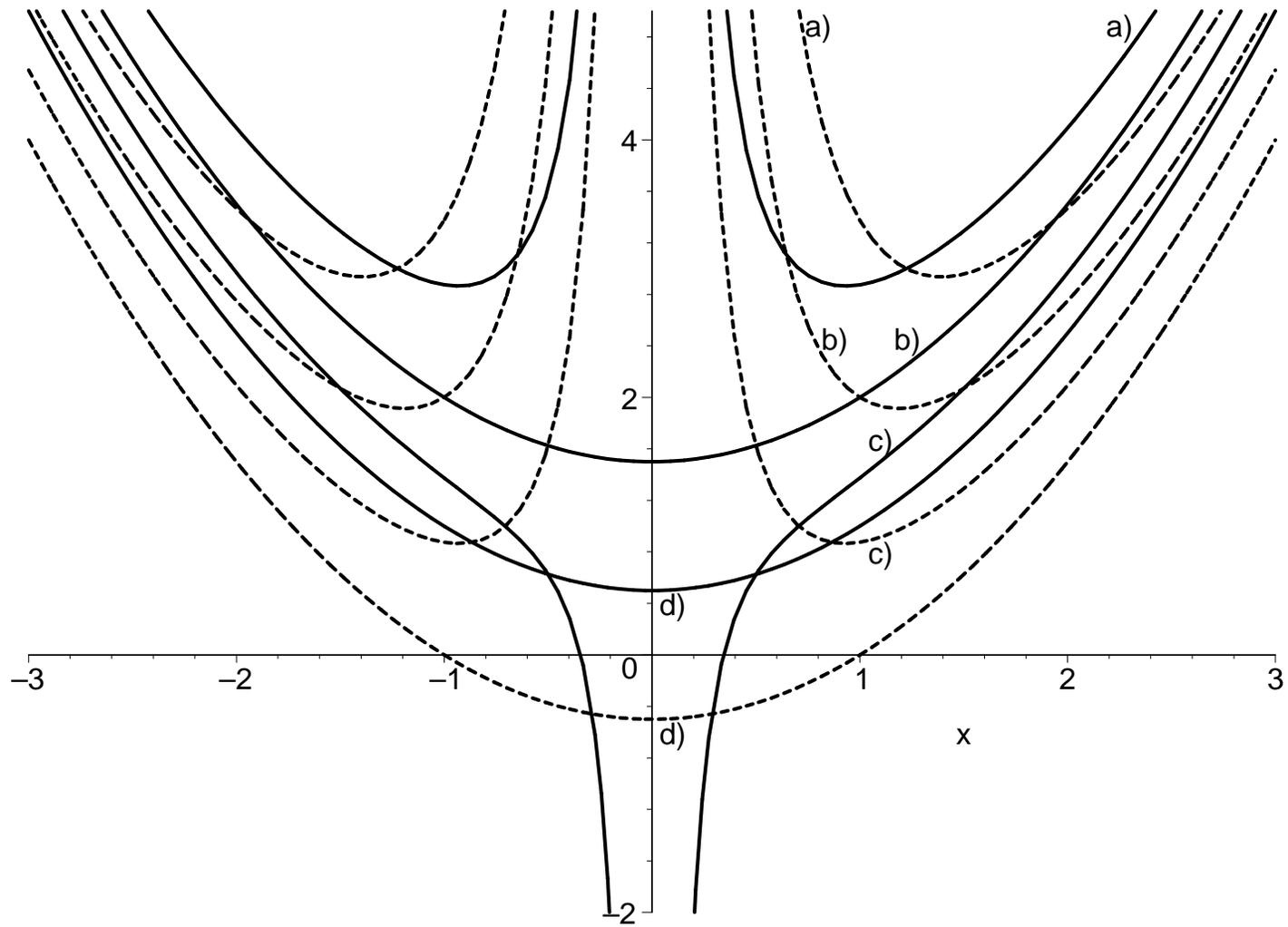
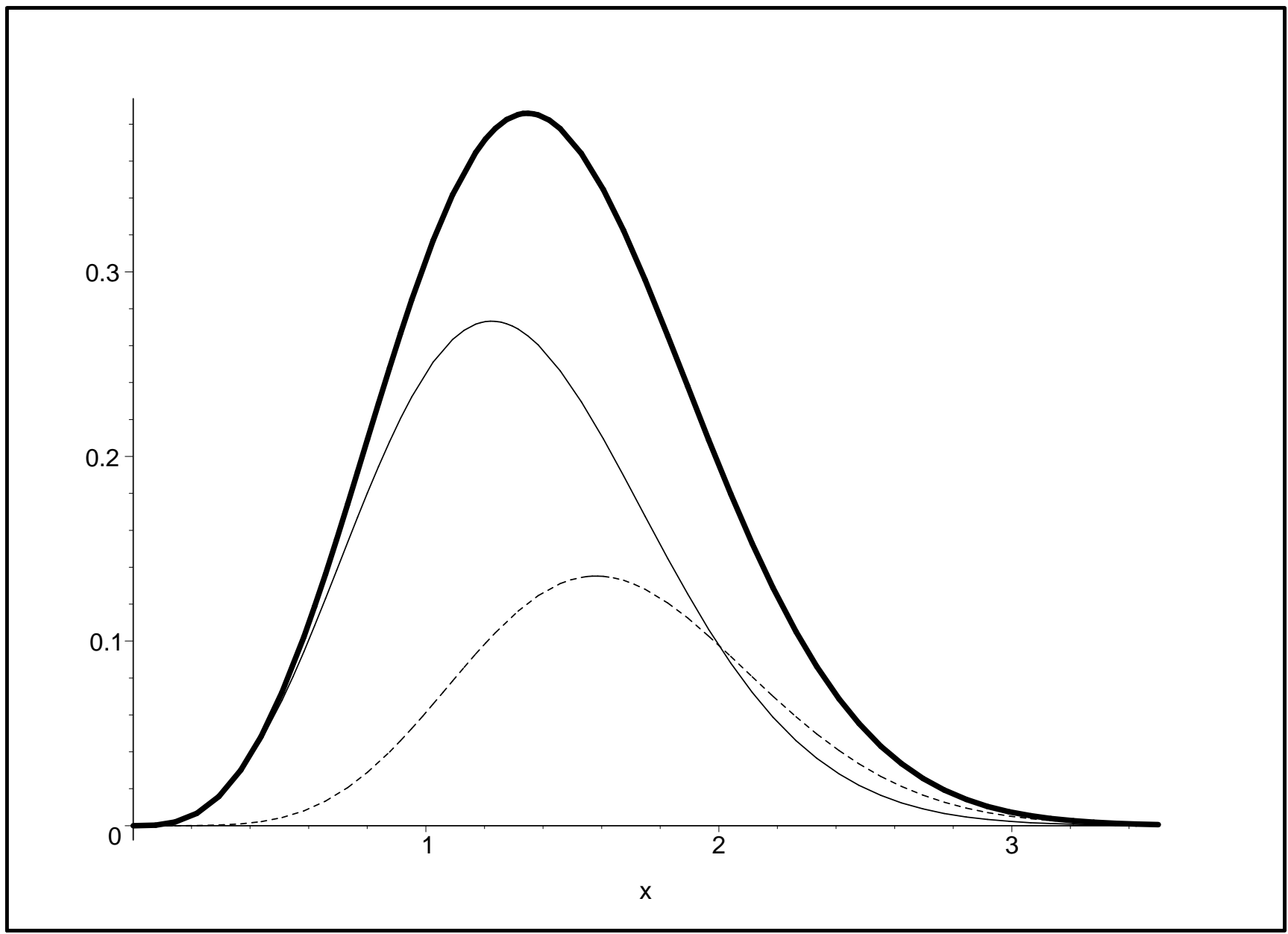
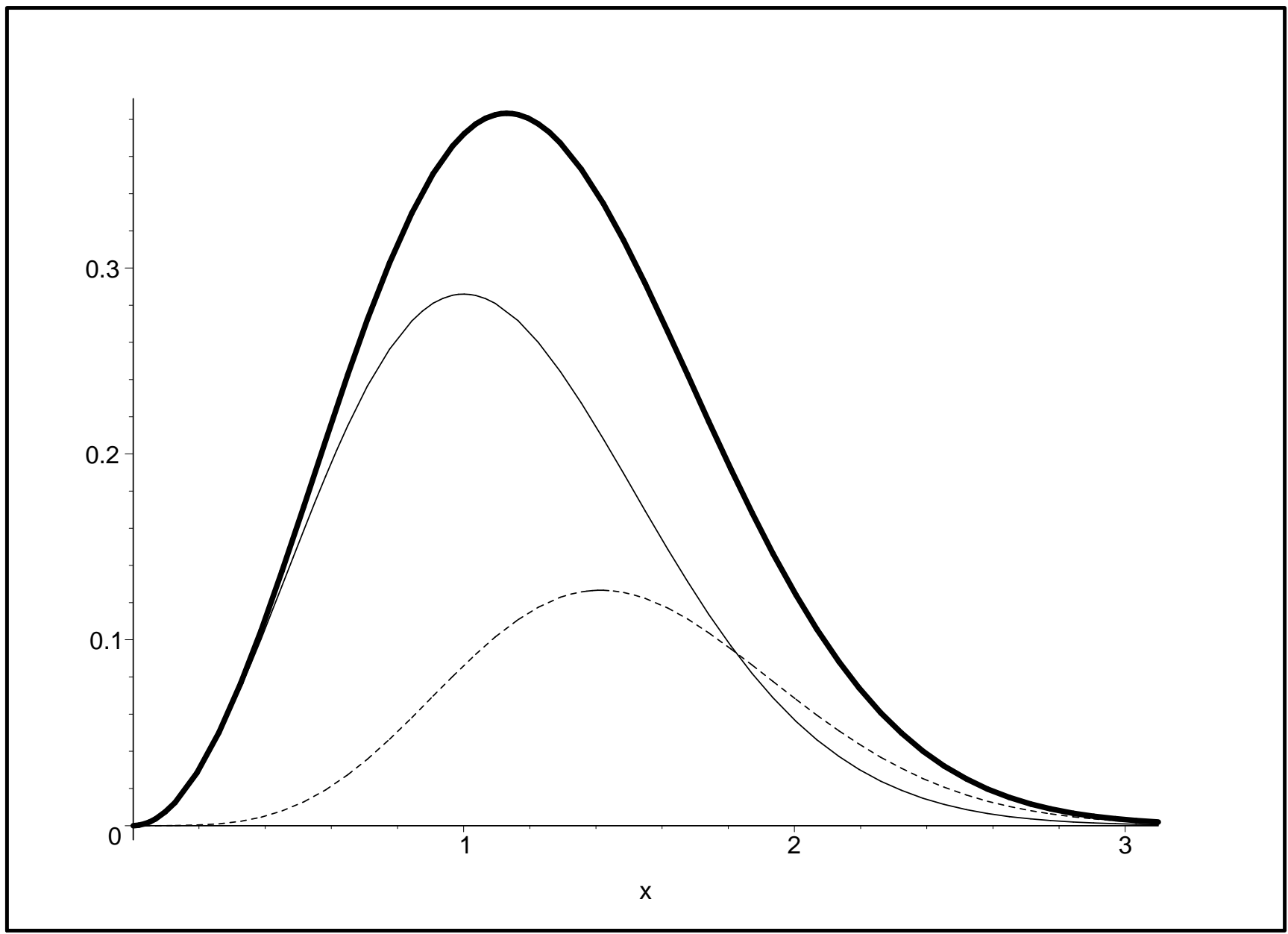
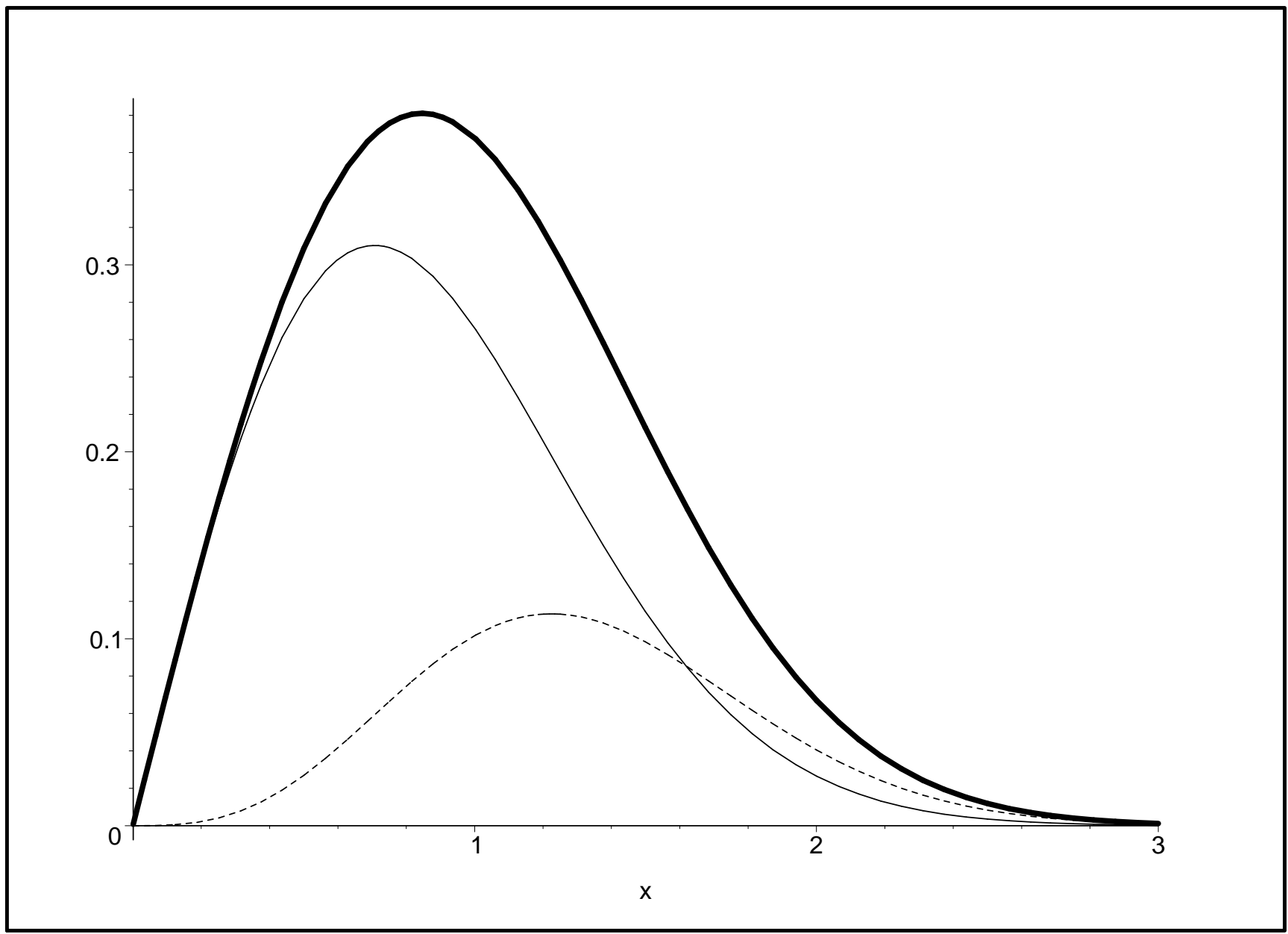
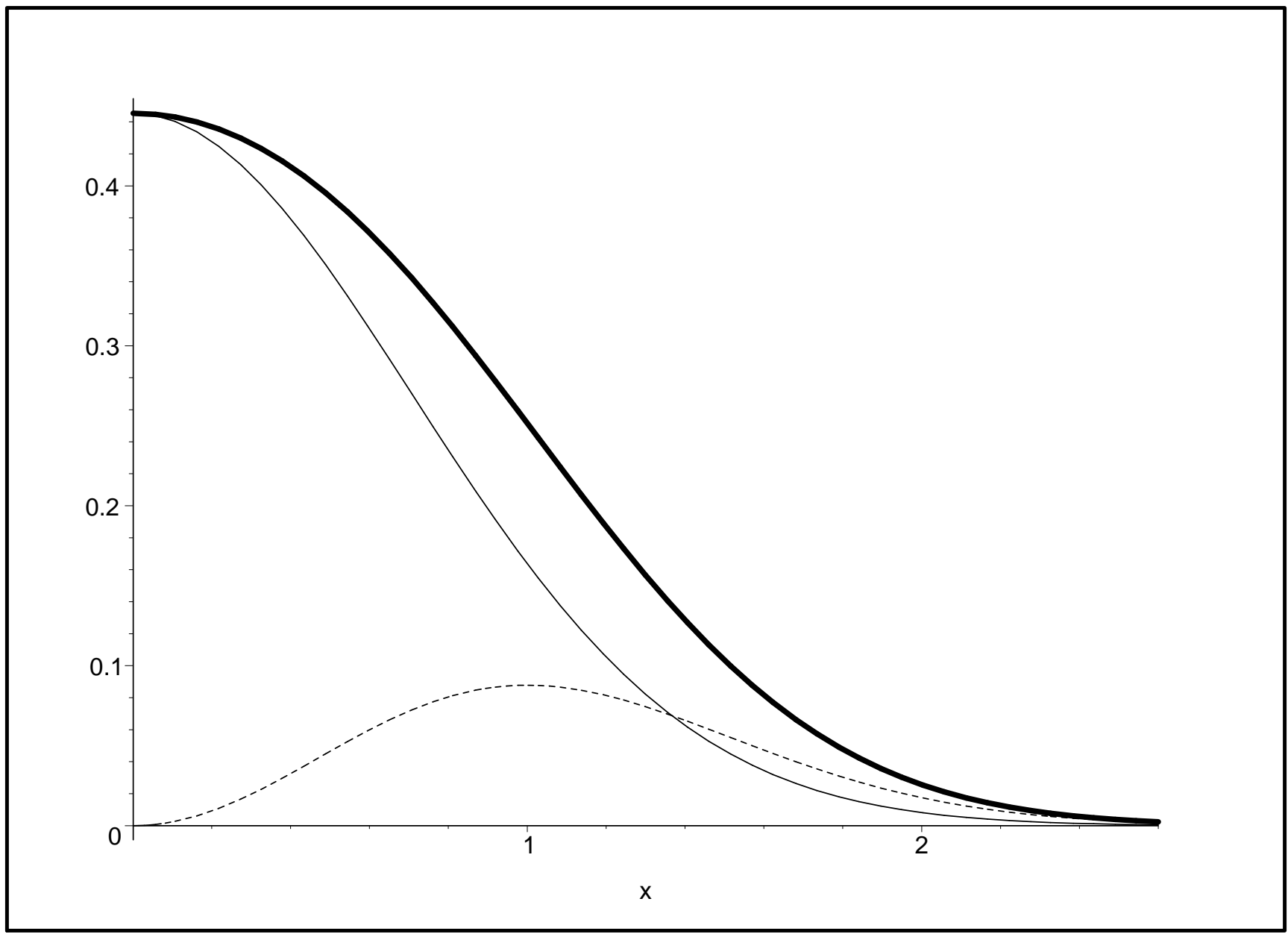
\end{document}